\documentclass[preprint]{aastex61}
\usepackage{amsmath,amstext}
\usepackage[T1]{fontenc}
\usepackage{apjfonts} 
\usepackage[shortcuts]{extdash}
\shorttitle{Transient in M51}
\shortauthors{Jencson et al.}

\begin{document}

\title{Discovery of an intermediate-luminosity red transient in M51 and its likely dust-obscured, infrared-variable progenitor}

\author{Jacob E.\ Jencson}
\affiliation{Division of Physics, Mathematics, and Astronomy, California Institute of Technology, Pasadena, CA 91125, USA}
\author{Scott M.\ Adams}
\affiliation{Division of Physics, Mathematics, and Astronomy, California Institute of Technology, Pasadena, CA 91125, USA}
\author{Howard E.\ Bond}
\affiliation{Department of Astronomy \& Astrophysics, Pennsylvania State University, University Park, PA 16802, USA}
\affiliation{Space Telescope Science Institute, 3700 San Martin Drive, Baltimore, MD 21218, USA}
\author{Schuyler D.\ van Dyk}
\affiliation{Caltech/IPAC, Mailcode 100-22, Pasadena, CA 91125, USA}
\author{Mansi M.\ Kasliwal}
\affiliation{Division of Physics, Mathematics, and Astronomy, California Institute of Technology, Pasadena, CA 91125, USA}
\author{John Bally}
\affiliation{Center for Astrophysics and Space Astronomy, University of Colorado, 389 UCB, Boulder, CO 80309, USA}
\author{Nadejda Blagorodnova}
\affiliation{Department of Astrophysics/IMAPP, Radboud University, Nijmegen, The Netherlands}
\author{Kishalay De}
\affiliation{Division of Physics, Mathematics, and Astronomy, California Institute of Technology, Pasadena, CA 91125, USA}
\author{Christoffer Fremling}
\affiliation{Division of Physics, Mathematics, and Astronomy, California Institute of Technology, Pasadena, CA 91125, USA}
\author{Yuhan Yao}
\affiliation{Division of Physics, Mathematics, and Astronomy, California Institute of Technology, Pasadena, CA 91125, USA}
\author{Andrew Fruchter}
\affiliation{Space Telescope Science Institute, 3700 San Martin Drive, Baltimore, MD 21218, USA}
\author{David Rubin}
\affiliation{Space Telescope Science Institute, 3700 San Martin Drive, Baltimore, MD 21218, USA}
\author{Cristina Barbarino}
\affiliation{The Oskar Klein Centre, Department of Astronomy, Stockholm University, AlbaNova, SE-106 91 Stockholm, Sweden}
\author{Jesper Sollerman}
\affiliation{The Oskar Klein Centre, Department of Astronomy, Stockholm University, AlbaNova, SE-106 91 Stockholm, Sweden}
\author{Adam A.\ Miller}
\affiliation{Center for Interdisciplinary Exploration and Research in Astrophysics (CIERA) and Department of Physics and Astronomy, Northwestern University, Evanston, IL 60208, USA}
\author{Erin K.\ S.\ Hicks}
\affiliation{Department of Physics \& Astronomy, University of Alaska Anchorage, 3211 Providence Drive, Anchorage, AK 99508, USA}
\author{Matthew A.\ Malkan}
\affiliation{University of California Los Angeles, Department of Astronomy \& Astrophysics, 430 Portola Plaza, Box 951547, Los Angeles, CA 90095-1547, US}
\author{Igor Andreoni}
\affiliation{Division of Physics, Mathematics, and Astronomy, California Institute of Technology, Pasadena, CA 91125, USA}
\author{Eric C.\ Bellm}
\affiliation{DIRAC Institute, Department of Astronomy, University of Washington, 3910 15th Avenue NE, Seattle, WA 98195, USA}
\author{Robert Buchheim}
\affiliation{Lost Gold Observatory, Society for Astronomical Sciences, Gold Canyon, AZ 85118, USA}
\author{Richard Dekany}
\affiliation{Caltech Optical Observatories, California Institute of Technology, Pasadena, CA 91125, USA}
\author{Michael Feeney}
\affiliation{Caltech Optical Observatories, California Institute of Technology, Pasadena, CA 91125, USA}
\author{Sara Frederick}
\affiliation{Department of Astronomy, University of Maryland, College Park, MD 20742, USA}
\author{Avishay Gal-Yam}
\affiliation{Department of Particle Physics and Astrophysics, Weizmann Institute of Science, Rehovot 76100, Israel}
\author{Robert D. Gehrz}
\affiliation{Minnesota Institute for Astrophysics, School of Physics and Astronomy, University of Minnesota, 116 Church Street, S. E., Minneapolis, MN 55455, USA}
\author{Matteo Giomi}
\affiliation{Humboldt-Universit\"{a}t zu Berlin, 
Newtonstra{\ss}e 15, D-12489 Berlin, Germany}
\author{Matthew J.\ Graham}
\affiliation{Division of Physics, Mathematics, and Astronomy, California Institute of Technology, Pasadena, CA 91125, USA} 
\author{Wayne Green}
\affiliation{Center for Astrophysics and Space Astronomy, University of Colorado, 389 UCB, Boulder, CO 80309, USA}
\author{David Hale}
\affiliation{Caltech Optical Observatories, California Institute of Technology, Pasadena, CA 91125, USA}
\author{Matthew J.\ Hankins}
\affiliation{Division of Physics, Mathematics, and Astronomy, California Institute of Technology, Pasadena, CA 91125, USA}
\author{Mark Hanson}
\affiliation{Stellar Winds Observatory, Dark Sky New Mexico, Hidalgo County, NM, USA}
\author{George Helou}
\affiliation{Caltech/IPAC, Mailcode 100-22, Pasadena, CA 91125, USA}
\author{Anna Y.\ Q.\ Ho}
\affiliation{Division of Physics, Mathematics, and Astronomy, California Institute of Technology, Pasadena, CA 91125, USA}
\author{T.\ Hung}
\affiliation{Department of Astronomy and Astrophysics, University of California, Santa Cruz, CA 95064, USA}
\author{Mario Juri\'c}
\affiliation{DIRAC Institute, Department of Astronomy, University of Washington, 3910 15th Avenue NE, Seattle, WA 98195, USA}
\author{Malhar R.\ Kendurkar}
\affiliation{Prince George Astronomical Observatory, 7765 Tedford Road, Prince George, BC V2N 6S2, Canada}
\author{S.\ R.\ Kulkarni}
\affiliation{Division of Physics, Mathematics, and Astronomy, California Institute of Technology, Pasadena, CA 91125, USA}
\author{Ryan M.\ Lau}
\affiliation{Institute of Space \& Astronautical Science, Japan Aerospace Exploration Agency, 3-1-1 Yoshinodai, Chuo-ku, Sagamihara,Kanagawa 252-5210, Japan}
\author{Frank J.\ Masci}
\affiliation{Caltech/IPAC, Mailcode 100-22, Pasadena, CA 91125, USA}
\author{James D.\ Neill}
\affiliation{Division of Physics, Mathematics, and Astronomy, California Institute of Technology, Pasadena, CA 91125, USA}
\author{Kevin Quin}
\affiliation{Northern Virginia Astronomy Club, Fairfax, VA 22030, USA}
\author{Reed L.\ Riddle}
\affiliation{Caltech Optical Observatories, California Institute of Technology, Pasadena, CA 91125, USA}
\author{Ben Rusholme}
\affiliation{Caltech/IPAC, Mailcode 100-22, Pasadena, CA 91125, USA}
\author{Forrest Sims}
\affiliation{Society for Astronomical Sciences, Desert Celestial Observatory, Gilbert, AZ 85233, USA}
\author{Nathan Smith}
\affiliation{University of Arizona, Steward Observatory, 933 N. Cherry Avenue, Tucson, AZ 85721, USA}
\author{Roger M.\ Smith}
\affiliation{Caltech Optical Observatories, California Institute of Technology, Pasadena, CA 91125, USA}
\author{Maayane T.\ Soumagnac}
%\author[0000-0001-6753-1488]{Maayane T.\ Soumagnac}
\affiliation{Benoziyo Center for Astrophysics, Weizmann Institute of Science, Rehovot, Israel}
\author{Yutaro Tachibana}
\affiliation{Department of Physics, Tokyo Institute of Technology, 2-12-1 Ookayama, Meguro-ku, Tokyo 152-8551, Japan}
\affiliation{Division of Physics, Mathematics, and Astronomy, California Institute of Technology, Pasadena, CA 91125, USA}
\author{Samaporn Tinyanont}
\affiliation{Division of Physics, Mathematics, and Astronomy, California Institute of Technology, Pasadena, CA 91125, USA}
\author{Richard Walters}
\affiliation{Caltech Optical Observatories, California Institute of Technology, Pasadena, CA 91125, USA}
\author{Stanley Watson}
\affiliation{Stellar Winds Observatory, Dark Sky New Mexico, Hidalgo County, NM, USA}
\author{Robert E.\ Williams}
\affiliation{Space Telescope Science Institute, 3700 San Martin Drive, Baltimore, MD 21218, USA}
\affiliation{Department of Astronomy and Astrophysics, University of California, Santa Cruz, CA 95064, USA}

\correspondingauthor{Jacob E. Jencson}
\email{jj@astro.caltech.edu}

\begin{abstract}
We present the discovery of an optical transient (OT) in Messier~51, designated M51~OT2019-1 (also ZTF\,19aadyppr, AT\,2019abn, ATLAS19bzl), by the Zwicky Transient Facility (ZTF). The OT rose over 15~days to an observed luminosity of $M_r=-13$ (${\nu}L_{\nu}=9\times10^6~L_{\odot}$), in the luminosity gap between novae and typical supernovae (SNe). Spectra during the outburst show a red continuum, Balmer emission with a velocity width of $\approx400$\,km\,s$^{-1}$, \ion{Ca}{2} and [\ion{Ca}{2}] emission, and absorption features characteristic of an F-type supergiant. The spectra and multiband light curves are similar to the so-called ``SN impostors'' and intermediate-luminosity red transients (ILRTs). We directly identify the likely progenitor in archival \textit{Spitzer Space Telescope} imaging with a $4.5~\mu$m luminosity of $M_{[4.5]}\approx-12.2$~mag and a $[3.6]-[4.5]$ color redder than $0.74$~mag, similar to those of the prototype ILRTs SN\,2008S and NGC\,300~OT2008-1. Intensive monitoring of M51 with \textit{Spitzer} further reveals evidence for variability of the progenitor candidate at [4.5] in the years before the OT. The progenitor is not detected in pre-outburst \textit{Hubble Space Telescope} optical and near-IR images. The optical colors during outburst combined with spectroscopic temperature constraints imply a higher reddening of $E(B-V)\approx0.7$~mag and higher intrinsic luminosity of $M_r\approx-14.9$~mag (${\nu}L_{\nu}=5.3\times10^7~L_{\odot}$) near peak than seen in previous ILRT candidates. Moreover, the extinction estimate is higher on the rise than on the plateau, suggestive of an extended phase of circumstellar dust destruction. These results, enabled by the early discovery of M51~OT2019-1 and extensive pre-outburst archival coverage, offer new clues about the debated origins of ILRTs and may challenge the hypothesis that they arise from the electron-capture induced collapse of extreme asymptotic giant branch stars.
\end{abstract}


\section{Introduction}\label{sec:intro}
Searches for transients in the nearby universe have uncovered a diverse array of hydrogen-rich stellar events occupying the luminosity range between that of novae and supernovae (SNe). As more well-characterized events have been found, some defined classes are beginning to emerge. Distinguishing among these classes, however, can be challenging as there is significant overlap in their observed properties. These include events often associated with luminous blue variables (LBVs; \citealp{humphreys94,smith06}) variably referred to as ``SN impostors,'' ``$\eta$~Carinae variables,'' or ``giant eruptions'' \citep{humphreys99,vandyk00,pastorello10,smith10,smith14}. The ``luminous red novae'' (LRNe) are believed to be associated with stellar mergers or common-envelope ejections, including the 1--3~$M_{\odot}$ contact-binary merger V1309 Sco \citep{tylenda11} and the B-type stellar merger V838 Mon 
\citep{bond03,sparks08}. Several extragalactic events are also suggested to be members of this class, e.g., M31 RV \citep{rich89,bond06,bond11}, M85~OT2006-1 \citep{kulkarni07}, NGC\,4490~OT2011-1 \citep{smith16}, and M101~OT2015-1 \citep{blagorodnova17}. 

We adopt the term ``intermediate-luminosity red transient'' (ILRT), originally suggested by \citet{bond09}, to refer to the class of SN impostors similar to two well-characterized prototypes: SN\,2008S \citep{prieto08b,botticella09,smith09} and the 2008 optical transient (OT) in NGC\,300 (NGC\,300~OT2008-1; \citealp{berger09,bond09,humphreys11}). As the name suggests, these events reach peak luminosities ($M_{r/R} \approx -13$ to $-14$) fainter than those of typical core-collapse SNe, but comparable to those of other impostors. The peak is followed by the monotonic decline of their optical light curves, and they also show reddened spectra suggestive of strong internal extinction from the immediate circumburst environment. Their spectra are similar to some LBV-related transients and LRNe, showing strong H, \ion{Ca}{2}, and rare [\ion{Ca}{2}] emission features superimposed on an absorption spectrum characteristic of an F-type supergiant. The intermediate-width H features indicate relatively low ejection velocities of a $\mathrm{few}\times100$~km~s$^{-1}$. Both NGC\,300~OT2008-1 and SN\,2008S had densely self-obscured progenitor stars detected in archival imaging with the \textit{Spitzer Space Telescope}, whose luminosities and inferred masses ($M \approx 9$--$15~M_{\odot}$; \citealp{prieto08a,prieto08b,thompson09,kochanek11}) are lower than those of classical LBVs. Both events have now faded below their pre-explosion luminosities in the IR \citep{adams16a}, suggesting that the explosions may have been terminal. Other proposed members of this class discovered in the past decade include PTF\,10fqs \citep{kasliwal11}, SN\,2010dn \citep{smith11}, and AT\,2017be \citep{cai18}, although their larger distances prevented the direct identification of progenitors.

Here, we present the discovery by the Zwicky Transient Facility (ZTF; \citealp{bellm19b,graham19}) of an OT in M51 with similar properties to those of previously observed ILRTs and other SN impostors. At a distance of only $8.6$~Mpc \citep{mcquinn16,mcquinn17}, this is the closest such event in over a decade, allowing us to identify and characterize a candidate progenitor star in archival \textit{Spitzer} images. In this Letter, we describe the discovery and early observations (Section~\ref{sec:disc}), our analysis of available archival imaging and identification of the likely progenitor (Section~\ref{sec:progenitor}), the early photometric evolution (Section~\ref{sec:phot}), and spectroscopic properties (Section~\ref{sec:spectra}). In Section~\ref{sec:discussion}, we discuss the properties of the OT and its progenitor in the context of similar transients, and suggest it is a member of the ILRT class offering new insights on their origins.

\section{Discovery and data collection}\label{sec:disc}
\subsection{Discovery in M51}
On UT 2019 January 22.6 (MJD = 58505.6), ZTF\,19aadyppr was detected as a new OT source in the nearby galaxy M51 by ZTF with the 48\,inch Samuel Oschin Telescope (P48) at Palomar Observatory as part of the public ZTF 3 day cadence survey of the visible Northern Sky. The detection passed significance \citep{masci19} and machine-learning thresholds \citep{tachibana18,mahabal19}, and was released as a public alert \citep{patterson19}. We refer to the phase, $t$, as the number of days since the first ZTF detection throughout this work. Located at an R.A. and decl. of $13^{\mathrm{h}}29^{\mathrm{m}}42\fs41, +47\degr11\arcmin16\farcs6$ (J2000.0), the source had an $r$-band AB magnitude at first detection of $19.6 \pm 0.2$. The source was not detected in an earlier image taken on 2019 January 19.6, to a limiting magnitude of $r > 20.5$, at $t=-3$~days. After a second detection on 2019 January 25.5 (MJD = 58508.5), ZTF\,19aadyppr was autonomously selected as a high-quality transient by the AMPEL analysis framework \citep{nordin19b}. The discovery was submitted to the Transient Name Server and provided the IAU designation AT\,2019abn \citep{nordin19a}. The source passed several ZTF science-program filters and was saved by human scanners for follow-up on the GROWTH Marshal \citep{kasliwal19}. Early spectroscopic observations reported by \citet{de19} on 2019 January 26 were characterized by a red continuum and strong, intermediate-width ($\approx 600$~km~s$^{-1}$) H$\alpha$ emission, consistent with a classification of SN impostor or young ILRT. An independent detection was reported to TNS on 2019 January 26 by the the ATLAS survey \citep{tonry18}, and the Las Cumbres Observatory (LCO) Global SN Project reported an additional spectrum taken on 2019 March 2 \citep{burke19} and ILRT classification. We use the name M51~OT2019-1 for this event hereafter.

\begin{figure}
\includegraphics[width=\textwidth]{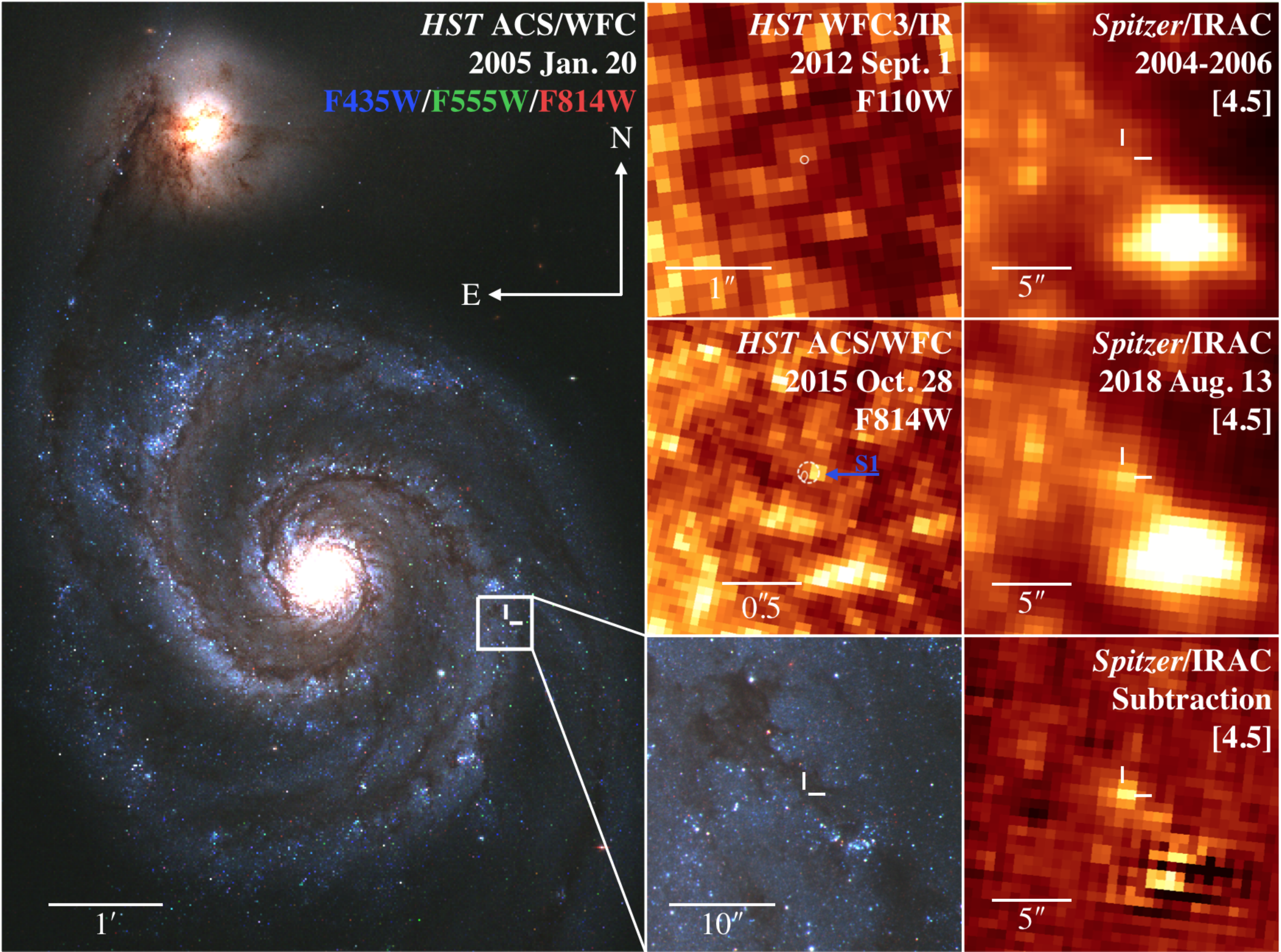}
\caption{\label{fig:images}
Pre-explosion \textit{HST} and \textit{Spitzer} imaging of M51~OT2019-1. In the leftmost panel, we show the color-composite \textit{HST} ACS/WFC mosaics of the M51 system from 2005 in three filters (F435W in blue, F555W in green, and F814W in red; PID: GO-10451; PI: S.\ Beckwith). The location of M51~OT2019-1 in a prominent dust lane along a spiral arm is indicated by the white cross-hairs and shown in more detail in the $30\arcsec \times 30\arcsec$ bottom center zoom-in panel. Above in the center column, we show the archival \textit{HST} coverage of the site in 2015 with ACS/WFC in F814W (center row) and in 2012 with WFC3/IR in F110W (top row). The 3-$\sigma$ error ellipses on the precise position of the transient from new \textit{HST}/WFC3 (solid) and Keck/NIRC2 (dashed) imaging are shown in white at the center of these panels. The nearest star-like object in the F814W image, labeled S1 in blue, is firmly outside the \textit{HST} error ellipse. In the rightmost column, we show the \textit{Spitzer}/IRAC archival [4.5] Super Mosaic (top), the most recent pre-explosion [4.5] image (center), and the subtraction of the two (bottom), clearly showing the variability of the coincident IR precursor source. 
}
\end{figure}

As shown in Figure~\ref{fig:images}, 
the transient was located in a star-forming spiral arm of M51, $108\farcs2$ from the galaxy's center. There is a prominent dust lane at the site, indicating the source may be subject to significant host extinction. Notably, M51 has a high rate of core-collapse SNe, with three known events discovered in the last 25\,yr: SN\,1994I (type~Ic; \citealp{schmidt94}), SN\,2005cs (type~II; \citealp{modjaz05}), and SN\,2011dh (type~IIb; \citealp{silverman11}). The NASA/IPAC Extragalactic Database\footnote{NED is operated by the Jet Propulsion Laboratory, California Institute of Technology, under contract with the National Aeronautics and Space Administration.} (NED) lists 50 individual distance measurements to M51, with a median value in distance modulus of 29.5~mag and a large standard deviation of $0.9$~mag. Throughout this work we assume a distance modulus for M51 from \citet{mcquinn16,mcquinn17} of $m-M = 29.67 \pm 0.02$ (statistical) $\pm 0.07$ (systematic; \citealp{rizzi07})~mag based on the luminosity of the tip of the red giant branch (TRGB) method, and that the systematic uncertainties associated with calibrating this method dominate over the statistical measurement uncertainties. We adopt the value from NED for the Galactic extinction toward M51 of $E(B-V) = 0.03$~mag, based on the \citet{schlafly11} recalibration of \citet{schlegel98}, and assuming a standard \citep{fitzpatrick99} reddening law with $R_V = 3.1$. 

\subsection{Imaging observations} \label{sec:imaging}
The field containing M51~OT2019-1 was regularly observed at many epochs with the ZTF camera on P48 in the $g$, $r$, and $i$ bands. The P48 images were reduced with the ZTF Science Data System pipelines \citep{masci19}, which perform image subtraction based on the \citet{zackay16}
algorithm and point-spread-function (PSF) photometry on the reference-subtracted images. We utilize this photometry for data from the public ZTF survey taken after 2019 January 1. For images taken as part of the ZTF Collaboration surveys and the Caltech surveys \citep{bellm19a}, and images taken prior to 2019 January 1 from the first ZTF Public Data Release\footnote{https://www.ztf.caltech.edu/page/dr1} (ZTF-DR1), we subsequently performed forced PSF-fitting photometry at the location of M51~OT2019-1, adopting a linear model for pixel values in the difference images as a function of the normalized PSF-model image values and use a Markov Chain Monte Carlo simulation to estimate the photometric uncertainties (Y.\ Yao et al.\ 2019, in preparation). Measurements for a given filter taken the same night were then averaged. 

Follow-up images in the $g^{\prime}$, $r^{\prime}$, $i^{\prime}$, and $Y$ bands were obtained with the Sinistro cameras on the Las Cumbres Observatory (LCO; \citealp{brown13}) 1\,m telescopes under the program NOAO2019A-011 (PI: N.~Blagorodnova). The data were reduced at LCO using the Beautiful Algorithms to Normalize Zillions of Astronomical Images (BANZAI) pipeline \citep{mccully18}. Photometry from the LCO $g^{\prime}r^{\prime}i^{\prime}$-band images were computed with the image-subtraction pipeline described in \citet{fremling16}, with template images from the Sloan Digital Sky Survey (SDSS; \citealp{ahn14}). The pipeline performs PSF-fitting photometry calibrated against several SDSS stars in the field. We performed aperture photometry on M51~OT2019-1 in the LCO $Y$-band, with the aperture radius set by the typical full width at half maximum (FWHM) of stars in the images, calibrated against several stars in the Pan-STARRS1 (PS1; \citealp{chambers16}) DR2 catalog \citep{flewelling16}.

Additional ground-based follow-up images were obtained in the optical and near-IR at several epochs using the Astrophysical Research Consortium Telescope Imaging Camera (ARCTIC; \citealp{huehnerhoff16}) and Near-Infrared Camera and Fabry-Perot Spectrometer (NICFPS; \citealp{vincent03}) on the Astrophysical Research Consortium (ARC) 3.5\,m Telescope at Apache Point Observatory (APO), and the Wide Field Infrared Camera (WIRC; \citealp{wilson03}) on the 200\,inch Hale Telescope (P200) at Palomar Observatory. Imaging of M51 at APO was obtained as part of a program to monitor 24 galaxies in the SPitzer Infrared Intensive Transients Survey (SPIRITS; PI: M.\ Kasliwal; PIDs 10136, 11063, 13053). Optical $g^{\prime}r^{\prime}i^{\prime}$ images were reduced in the standard fashion using bias, dark, and twilight flat-field frames. Our near-IR $JHK_{s}$ imaging employed large dithers alternating between the target and a blank sky field approximately every minute to allow for accurate subtraction of the bright near-IR sky background, and individual frames were flat-fielded, background-subtracted, astrometrically aligned with a catalog of Two Micron All Sky Survey (2MASS; \citealp{skrutskie06}) sources, and stacked. We performed aperture photometry at the location of the transient, and used the $g^{\prime}r^{\prime}i^{\prime}$ or $JHK_s$ magnitudes of several isolated stars in SDSS or 2MASS to measure the photometric zero-points in the optical and near-IR images, respectively. 

We also present photometry from CCD images obtained by F.\ Sims, R.\ Buchheim, W.\ Green, M.\ Hanson, and S.\ Watson, in the $V$ and $R$-bands, and by K.\ Quin and M.\ Kendurkar in the Astrodon LRGB filters\footnote{Astrodon filter information is available here: \url{https://astrodon.com/products/astrodon-lrgb-gen2-i-series-tru-balance-filters/}}. For the LRGB frames, we performed aperture photometry on the OT calibrated to standard magnitudes in $R$, $R$, $V$, and $B$, respectively, using SDSS stars in the field and adopting the conversions of \citet{jordi06}. 
Observations were obtained in the 3.6 and 4.5~$\mu$m imaging channels, [3.6] and [4.5], of the Infrared Array Camera (IRAC; \citealp{fazio04}) on board \textit{Spitzer} \citep{werner04,gehrz07} on 2019 April 7.6 as part of regular monitoring of M51 as part of SPIRITS. We performed image subtraction and aperture photometry as described in more detail for the archival \textit{Spitzer}/IRAC imaging in Section~\ref{sec:prog_var}.

\begin{figure}
\includegraphics[width=0.49\textwidth]{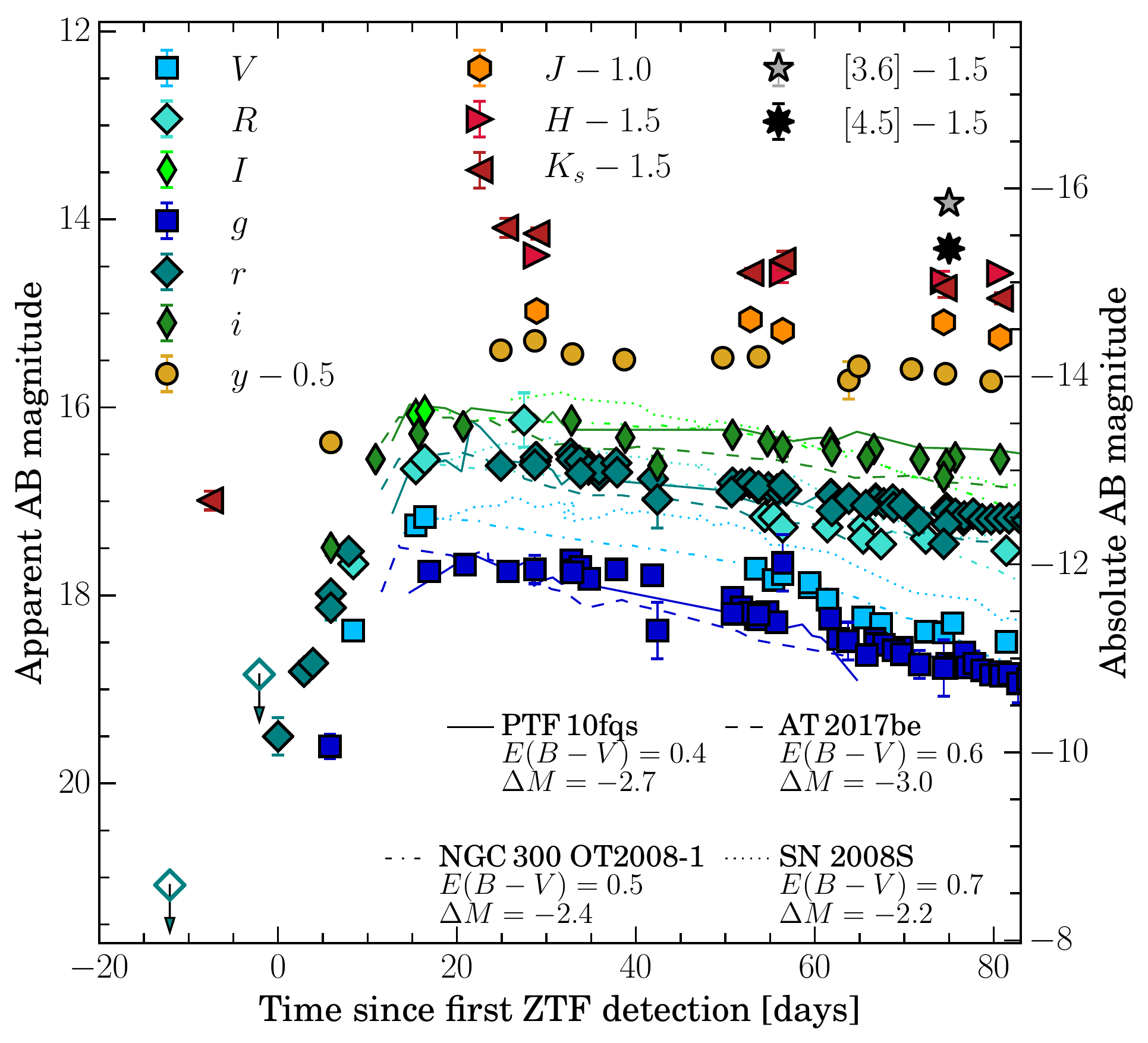}
\includegraphics[width=0.49\textwidth]{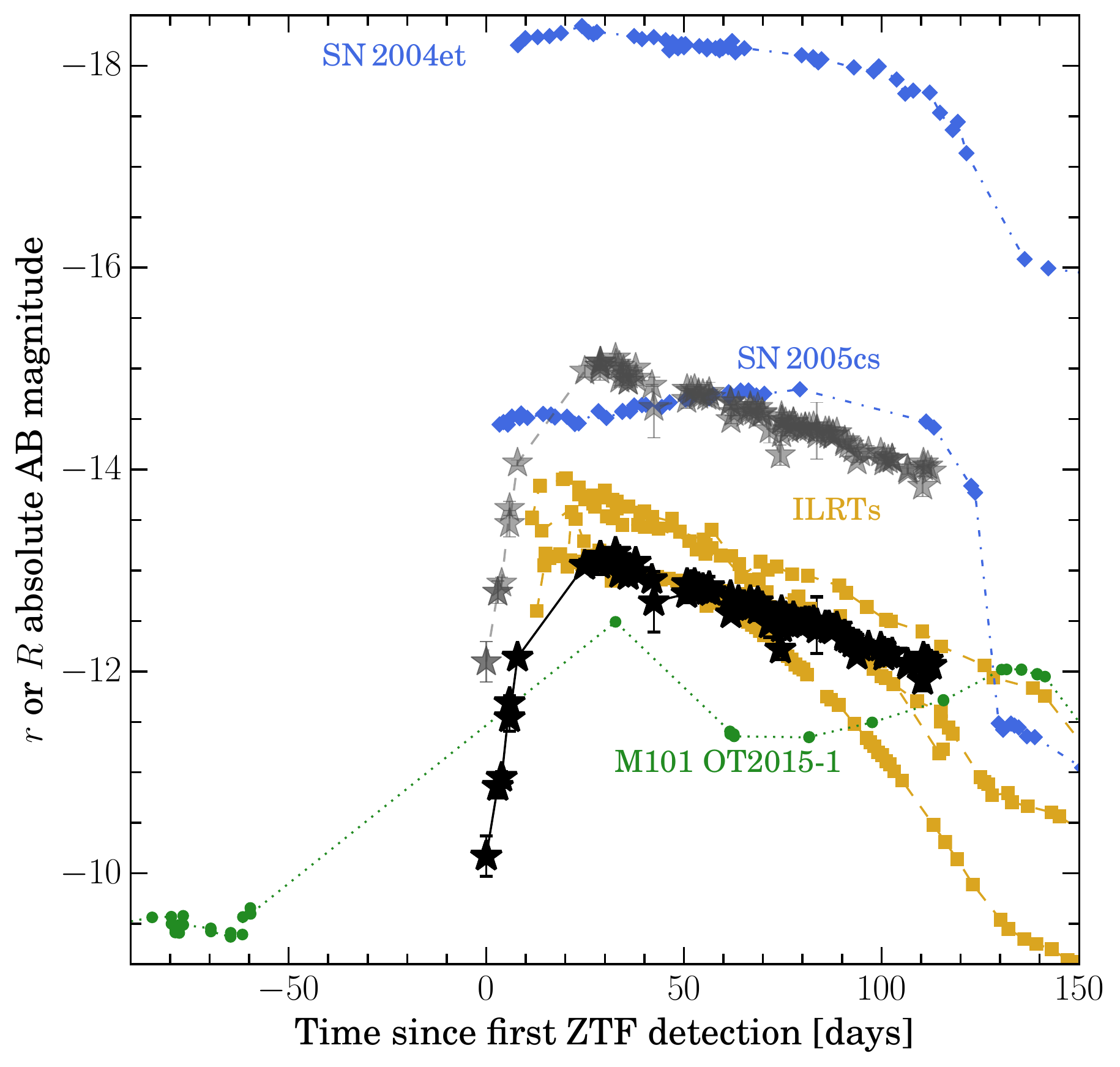}
%\plottwo{M51OT_lcs.pdf}{M51OT_lcs_r_compare.pdf}
\caption{\label{fig:lcs}
Left: multiband light curves of M51~OT2019-1 are shown as large filled symbols, and unfilled points with downward arrows represent upper limits from nondetections. Time on the $x$-axis is measured in days since the first detection by ZTF on 2019 January 22.6. We also show the $VRI$ measurements reported by \citet{pessev19}. Measurements have been corrected for Galactic extinction to M51 only, and $VRIJHK_s$ measurements have been converted from Johnson--Cousins/2MASS Vega magnitudes to AB magnitudes adopting \citet{blanton07} conversions. Absolute magnitudes at the assumed distance to M51 and assuming no host extinction are given on the $y$-axis to the right. For comparison, we show $gri$ light curves of PTF10fqs (solid lines; \citealp{kasliwal11}), and the $VRI$ light curves of SN\,2008S (dotted lines; \citealp{botticella09}) and NGC\,300~OT2008-1 (dashed--dotted lines, \citealp{humphreys11}). The comparison light curves have been corrected for Galactic extinction to their respective hosts from NED, then reddened to match the $g-r$ or $V-R$ colors of M51~OT2019-1 on the plateau and offset in absolute magnitude by the $\Delta M$ values indicated on the figure to match the vertical level of the M51~OT2019-1 light curves. Right: $r$-band light curves of M51 OT2019-1 are shown as the black stars, and corrected for our estimate of the total host/CSM extinction near peak with $E(B - V) = 0.7$~mag as lighter gray stars. The $r$ or $R$ light curves of the comparison ILRTs are shown as yellow squares, also corrected with an estimate for the total extinction near peak as described in the text. We compare to the $r$- or $R$-band light curves of other hydrogen-rich transients including the Type IIP SN\,2004et \citep{maguire10}, low-luminosity Type IIP SN\,2005cs \citep{pastorello06}, and the LRN M101~OT2015-1 \citep{blagorodnova17}.
}
\end{figure}

Our photometry of M51~OT2019-1 is shown in Figure~\ref{fig:lcs} on the AB magnitude system, corrected for Galactic extinction. Where appropriate we have converted $VRIJHK_s$ magnitudes on the Vega system to AB magnitudes using the conversions of \citet{blanton07}.

We also obtained high-resolution, adaptive-optics $J$-band imaging of the transient on 2019 February 16.6, using the near-IR camera (NIRC2; PI: K.\ Matthews) on the 10\,m Keck~II Telescope on Maunakea. Further high-resolution imaging was obtained on 2019 March 5.0 with the \textit{Hubble Space Telescope} (\textit{HST}) WFC3 camera in the F275W, F336W, and F814W filters as the first target in a test program proposing a new method of \textit{HST} observing (PI: A.\ Fruchter; PID SNAP-15675). This new method, called ``Rolling Snapshots,'' allows fairly rapid response by updating a list of snapshot targets weekly; however, due to present limitations in the Astronomer's Proposal Tool (APT), the observing is done in a custom subarray mode.  Data were reduced using the standard pipelines at STScI, adding a charge-transfer-efficiency (CTE) correction (which was not done by the pipeline due to the use of the custom subarray). CTE-corrected frames were then combined into image mosaics for each filter using AstroDrizzle within PyRAF to attempt to flag cosmic-ray hits. We performed PSF-fitting photometry for the transient using DOLPHOT \citep{dolphin00,dolphin16} and obtained $23.32 \pm 0.17$, $20.47 \pm 0.03$, and $15.754 \pm 0.002$~mag (Vega scale) for F275W, F336W, and F814W, respectively. 

\subsection{Spectroscopic observations}
%%rewriting to make more concise!! 
We obtained a sequence of optical spectra of M51~OT2019-1 using several instruments covering phases from $t=4$ to $111$\,days. This includes four spectra with the Alhambra Faint Object Spectrograph and Camera (ALFOSC) on the 2.56\,m Nordic Optical Telescope (NOT) at the Spanish Observatorio del Roque de los Muchachos on La Palma, including two low-resolution spectra with the 300\,lines\,mm$^{-1}$ grism (Gr4) and an epoch of intermediate-resolution spectra using two 600\,lines\,mm$^{-1}$ grisms (Gr7 and Gr8), five spectra with the Double Beam Spectrograph (DBSP; \citealt{Oke1982}) on P200, one spectrum with the Gemini Multi-Object Spectrograph (GMOS; \citealt{Hook2004}) on the Gemini North Telescope through our Target of Opportunity program (PI: A.\ Miller; PID GN-2018B-Q-132), one spectrum with the Spectral Energy Distribution Machine (SEDM; \citealp{blagorodnova18}) on the Palomar 60\,inch Telescope (P60), and one spectrum with the Low Resolution Imaging Spectrometer (LRIS; \citealp{goodrich03}) on the Keck I Telescope. The spectra were reduced using standard techniques including wavelength calibration with arc-lamp spectra and flux calibration using spectrophotometric standard stars. In particular, we made use of a custom PyRAF-based reduction pipeline\footnote{\url{https://github.com/ebellm/pyraf-dbsp}} \citep{bellm16} for DBSP spectra, the IDL-based reduction and pipeline LPipe\footnote{\url{http://www.astro.caltech.edu/~dperley/programs/lpipe.html}} \citep{perley19} for the LRIS spectrum, the fully automated Python-based reduction pipeline pysedm\footnote{\url{https://github.com/MickaelRigault/pysedm}} \citep{rigault19} for the SEDM spectra, and standard tasks in Gemini IRAF package\footnote{\url{http://www.gemini.edu/sciops/data-and-results/processing-software}} for the GMOS spectrum following procedures provided in the GMOS Data Reduction Cookbook\footnote{\url{http://ast.noao.edu/sites/default/files/GMOS_Cookbook/}}. 

We also obtained an epoch of near-IR spectroscopy of M51~OT2019-1 with the TripleSpec spectrograph \citep{herter08} on P200. We obtained four exposures of the transient (300~s each) while nodding the transient along the slit between exposures to allow sky subtraction. The data were reduced with a modified version of the IDL-based data reduction package Spextool\footnote{\url{http://irtfweb.ifa.hawaii.edu/~cushing/spextool.html}} \citep{cushing04} for P200/TripleSpec. Corrections for the strong near-IR telluric absorption features and flux calibrations were performed with observations of the A0\,V standard star HIP~61471, using the method developed by \citet{vacca03} implemented in the IDL tool xtellcor as part of Spextool. 

\begin{figure}
\includegraphics[width=\textwidth]{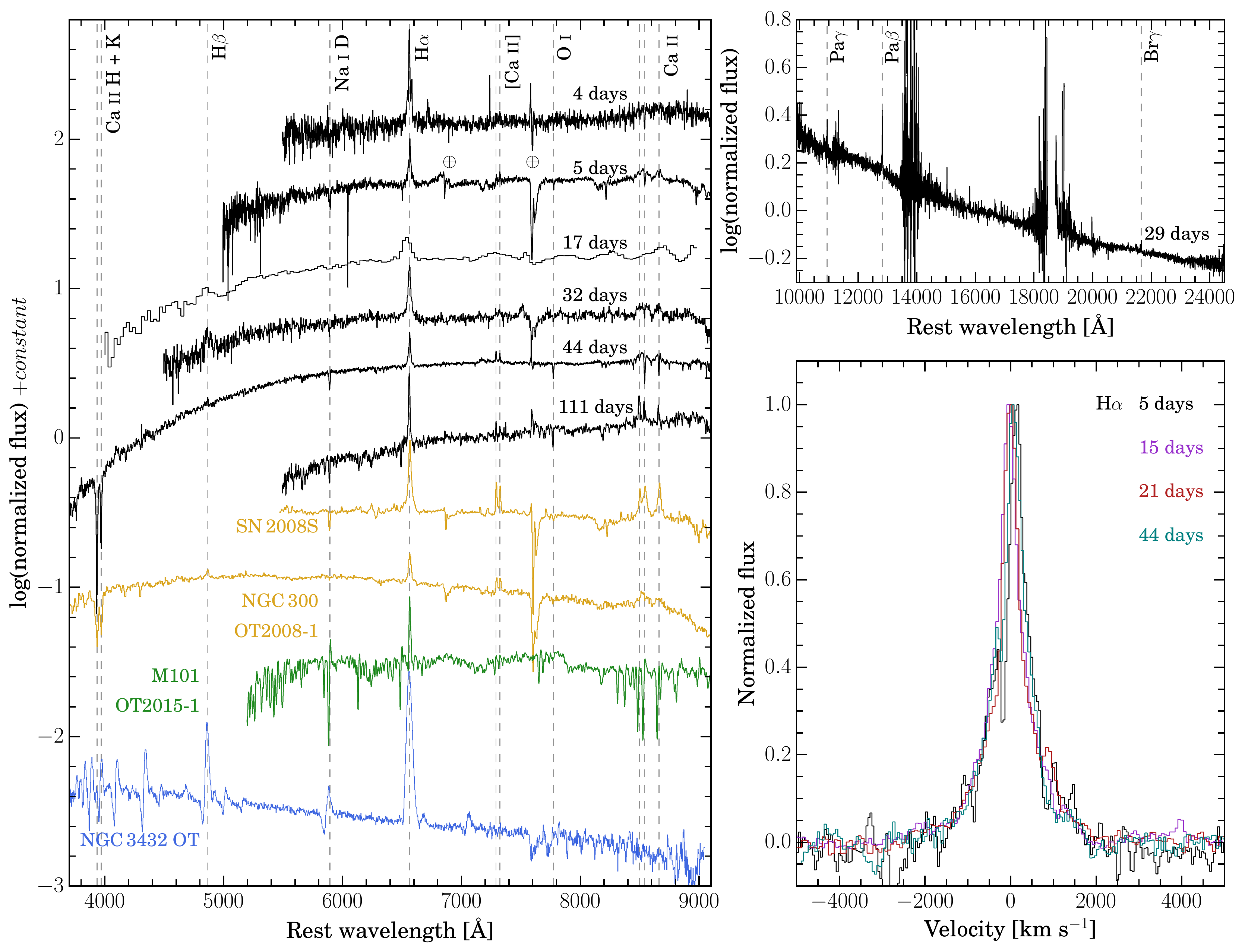}
\caption{\label{fig:spec}
In the left panel, we show the optical spectral evolution of M51~OT2019-1 in black with the phase of each spectrum indicated along the right side of the panel. We show spectra of other ILRTs in yellow (SN\,2008S from \citealp{botticella09} and NGC\,300~OT2008-1 from \citealp{bond09}), the M101 LRN \citep{blagorodnova17} in green, and the 2008 LBV outburst in NGC\,3432 \citep{pastorello10} in blue for comparison. Each spectrum has been normalized to the continuum flux level around 7000~\AA, and shifted vertically by an arbitrary constant for clarity. In the top right panel, we show the near-IR spectrum of M51~OT2019-1 from $t=29$~days normalized to the continuum level around 16000~\AA. We indicate the locations of several spectral features with gray, dashed vertical lines as labeled along the top of each panel. In the bottom right panel, we show the continuum-subtracted H$\alpha$ velocity profiles for several epochs in the early evolution of the transient. The apparent absorption component in the $t=5$~days spectrum is an artifact of the background subtraction. 
}
\end{figure}

A complete log of our spectroscopic observations is provided in Table~\ref{table:spec}, and a representative set of our spectral sequence is shown in Figure~\ref{fig:spec}. All of our spectra will be made publicly available at the Weizmann Interactive Supernova Data
Repository\footnote{\url{https://wiserep.weizmann.ac.il}} (WISeREP; \citealp{yaron12}). 

\begin{deluxetable*}{lcrlcc}
\tablecaption{Log of spectroscopic observations \label{table:spec}}
\tablehead{\colhead{UT Date} & \colhead{MJD} & \colhead{Phase} & \colhead{Tel./Instr.} & \colhead{Range} & \colhead{Resolution} \\ 
\colhead{} & \colhead{} & \colhead{(days)} & \colhead{} & \colhead{(\AA)} & \colhead{($\lambda / \delta \lambda$)} }
\startdata
2019 Jan 26 & 58509 & 4   & NOT/ALFOSC      & 4000--9000   & 280  \\ %1.3" slit
2019 Jan 26 & 58509 & 4   & P200/DBSP       & 3500--10000  & 700  \\ %2.0" slit
2019 Jan 27 & 58510 & 5   & Gemini N/GMOS   & 5000--10000  & 1000 \\ %1" slit
2019 Feb 6  & 58520 & 15  & NOT/ALFOSC      & 3650--7110   & 500  \\ %slit width?, assuming 1.3"
2019 Feb 6  & 58520 & 15  & NOT/ALFOSC      & 5680--8580   & 700  \\ %slit width?, assuming 1.3"
2019 Feb 8  & 58522 & 17  & P60/SEDM        & 3650--10000  & 100  \\
2019 Feb 12 & 58526 & 21  & P200/DBSP       & 3500--10000  & 1000 \\ %1.5" slit
2019 Feb 20 & 58534 & 29  & P200/TripleSpec & 10000--24000 & 2600 \\ %
2019 Feb 23 & 58537 & 32  & NOT/ALFOSC      & 3200--9600   & 280  \\ %slit width?, assuming 1.3"
2019 Mar 7  & 58549 & 44  & Keck I/LRIS     & 3500--10300  & 600/1000\tablenotemark{a} \\
2019 Mar 16 & 58558 & 53  & P200/DBSP       & 3500--10000  & 1000 \\ %1.5" slit
2019 Apr 13 & 58586 & 81  & P200/DBSP       & 3500--10000  & 1000 \\ %1.5" slit
2019 May 13 & 58616 & 111 & P200/DBSP       & 3500--10000  & 1000 \\ %1.5" slit
\enddata
\tablenotetext{a}{Values given for the blue and red sides of LRIS, respectively.}
\end{deluxetable*}

\section{Analysis}\label{sec:analysis}

\subsection{Archival imaging and progenitor constraints}\label{sec:progenitor}

\subsubsection{Progenitor candidate identification}\label{sec:prog_id}

We searched for the presence of a progenitor star in archival imaging taken with \textit{HST} and \textit{Spitzer}/IRAC. To determine the precise position of the transient in the archival \textit{HST} imaging, we registered the NIRC2 $J$-band image of the transient with the 2015 ACS/WFC F814W image (PID: GO-13804; PI: K.\ McQuinn) and 2012 WFC3/IR F110W image (PID: 12490; PI: J.\ Koda). Using several stars and compact background galaxies in common between the new and archival frames, we achieved rms astrometric uncertainties on the position of the transient of 0.44 ACS pixels ($0\farcs022$) in the F814W image and 0.12 WFC3/IR pixels ($0\farcs015$) in the F110W image. We identified an apparent point source at the edge of our 3$\sigma$ error circle in the F814W image (object S1 in Figure~\ref{fig:images}), but the precision of our registration was insufficient to establish or rule out coincidence with the transient. We thus triggered the new \textit{HST} observations with WFC3/UVIS, described above in Section~\ref{sec:imaging}, in order to obtain a more precise position. 

We repeated the registrations as above, now using the new WFC3/UVIS F814W image and obtained improved rms astrometric uncertainties on the $(x,y)$ position of the transient of $(0.13,0.20)$ ACS pixels or $(0\farcs007,0\farcs01)$ in the archival F814W image and 0.1 WFC3/IR pixels ($0\farcs01$) in the F110W image. We show the 3$\sigma$ error ellipses on the position of the transient in each of these images in Figure~\ref{fig:images}. We performed PSF-fitting photometry on the archival \textit{HST} images used for registration using DOLPHOT. The nearest star detected in the 2015 F814W frame (object S1; $0\farcs06$ from the transient location) is firmly outside the 3$\sigma$ error ellipse. We note that there is an apparent source at the location of the transient in the archival F110W images, but it appears spatially extended and was not detected by DOLPHOT. Furthermore, the precise position of the transient is offset from the apparent centroid of this source. We thus find it is unlikely that the emission at the location is primarily due to the progenitor, and is more likely a blend of nearby, contaminating sources. 

\begin{figure}
\plotone{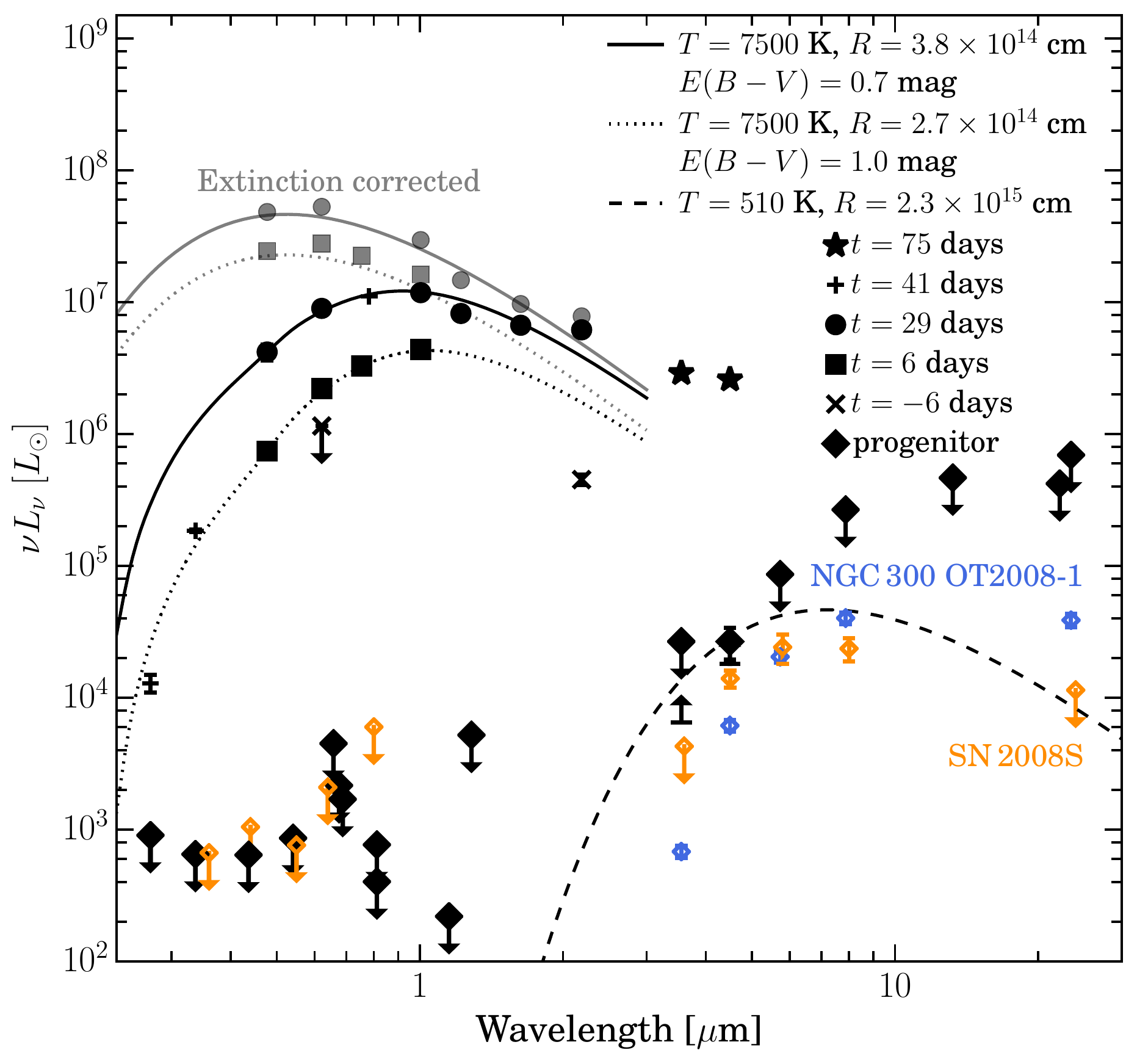}
\caption{\label{fig:sed} SEDs from photometry are shown in black for the M51\,OT2019-1 progenitor (diamonds), and at multiple phases in the evolution of the transient ($t=-$6, 6, 29, 41, and 75~days as crosses, squares, circles, plusses, and stars, respectively). Upper limits from nondetections are indicated with downward arrows. Lower limits on the progenitor SED from image subtraction are also indicated by upward arrows in black. Blackbody approximations to the data are shown for the progenitor (dashed curve), $t=6$~days SED on the rise (dotted) and $t=29$~days SED on the plateau with radii and temperatures given in the legend. For post-discovery SEDs of the transient, the curves shown have been reddened by the amount listed. The corresponding dereddened data and blackbody curves are shown in gray. For comparison, the progenitor SED data are shown for NGC\,300~OT2008-1 (open blue diamonds; \citealp{prieto08a}) and SN\,2008S (open orange diamonds; \citealp{prieto08b}).
}
\end{figure}

We then used DOLPHOT to obtain limits on the progenitor flux in the extensive archival coverage of the site with \textit{HST}, including the images from 2005 with ACS/WFC in F435W, F555W, F814W, and F658N (PID: GO-10452; PI: S.\ Beckwith), 2012 with WFC3/UVIS in F689M and F673N (PID: GO-12762; PI: K.\ Kuntz), 2012 with WFC3/IR in F110W and F128N (PID: GO-12490; PI: J.\ Koda), 2014 with WFC3/UVIS in F275W and F336W (PID: GO-13340; PI: S.\ van Dyk), and 2015 with ACS/WFC in F606W and F814W (PID: GO-13804, PI: K.\ McQuinn). We adopt 5$\sigma$ limiting magnitudes based on the detection significance of the faintest sources from DOLPHOT within a 100\,pixel radius of the transient position. Our full set limits on the progenitor from \textit{HST} are shown in Figure~\ref{fig:sed}, converted to band luminosities ($\nu L_{\nu}$) at the assumed distance to M51 and correcting for Galactic extinction. 

Utilizing the extensive archival coverage in the four imaging channels of \textit{Spitzer}/IRAC, we created deep stacks of all available images and registered them to the new \textit{HST} imaging with an astrometric rms of $0\farcs4$. A source consistent with the position of M51~OT2019-1 ($0\farcs6$ away) is clearly detected in the [4.5] stack. We do not see a clear point source at the location in the other channel stacks. Based on aperture photometry the source has a [4.5] flux of $0.0177\pm0.0048$ mJy and 3-$\sigma$ limiting fluxes of $<0.014$, $<0.073$, and $<0.31$~mJy in [3.6], [5.8], and [8.0], respectively. Given the high degree of crowding/blending with nearby emission at [4.5], our measurement of the flux of the putative progenitor may suffer contamination, but as we argue in more detail below and in Section~\ref{sec:prog_var}, the majority of the flux is likely attributable to a single source.

We also examined archival imaging coverage of the site with the Multiband Imaging Photometer for \textit{Spitzer} (MIPS) including 24~$\mu$m Super Mosaic (2004--2008 stack), and the 70 and 160~$\mu$m images from 2004 (PID: 159; PI: R.\ Kennicutt). We derive 3-$\sigma$ limiting fluxes from aperture photometry of $<2.4$, $<49.0$, and $350.0$~mJy in the three MIPS imaging bands, respectively. Similarly, we derived limits on the flux of the progenitor from the available imaging coverage with the \textit{Wide-field Infrared Survey Explorer} (\textit{WISE}; \citealp{wright10}) from the AllWISE Image Atlas\footnote{Images are available here: \url{https://irsa.ipac.caltech.edu/applications/wise/}} at 3.4, 4.6, 12, and 22~$\mu$m. Our limits of $0.20$, $0.41$, $0.91$, and $1.4$~mJy in the four \textit{WISE} channels, respectively, are typically less constraining than the corresponding limits from \textit{Spitzer} largely due to the coarser spatial resolution in \textit{WISE} imaging.

Using zero-magnitude fluxes (Vega system) given in the \textit{Spitzer}/IRAC Handbook\footnote{\url{https://irsa.ipac.caltech.edu/data/SPITZER/docs/irac/iracinstrumenthandbook/}}, we find the precursor source to have $M_{[4.5]} = -12.2$~mag and $[3.6] - [4.5] > 0.74$~mag. Sources with these IR properties are exceptionally rare \citep{thompson09,khan10}, indicating that the [4.5] emission is likely associated with a single star, and furthermore the spatial coincidence with M51~OT2019-1 strongly suggests a physical association. The spectral energy distribution (SED) is markedly similar to those of the self-obscured progenitors of SN\,2008S and NGC\,300~OT2008-1 (see Figure~\ref{fig:sed}). The best-fit blackbody, incorporating the [4.5] detection and [3.6], [5.8], and [8.0] limits, has a temperature of $T \approx 510$~K and luminosity of $L \approx 6.3 \times 10^4~L_{\odot}$, though there is significant uncertainty in these parameters. Our constraints at wavelengths $> 10~\mu$m are also relatively weak. Thus, we are unable to rule out emission from cooler circumstellar dust that would indicate a more luminous (and hence more massive) progenitor. 

\subsubsection{Pre-explosion variability}\label{sec:prog_var}
We examined the considerable coverage of the location with \textit{Spitzer}/IRAC at [3.6] and [4.5] for historical variability of the progenitor, including regular monitoring since 2014 as part of SPIRITS. The post-basic calibrated data (PBCD) level images were downloaded from the \textit{Spitzer} Heritage Archive\footnote{\url{https://sha.ipac.caltech.edu/applications/Spitzer/SHA/}} and \textit{Spitzer} Early Release Data Service\footnote{\url{http://ssc.spitzer.caltech.edu/warmmission/sus/mlist/archive/2015/msg007.txt}} and processed through an automated image-subtraction pipeline (for details see \citealp{kasliwal17}). For reference images, we used the Super Mosaics\footnote{Super Mosaics are available as \textit{Spitzer} Enhanced Imaging Products through the NASA/IPAC Infrared Science Archive: \url{https://irsa.ipac.caltech.edu/data/SPITZER/Enhanced/SEIP/overview.html}} consisting of stacks of images obtained between 2004 May 18 and 2006 January 29. In addition to the single epoch images, we performed image subtraction using the same references on our deep [3.6] and [4.5] stacks of all available pre-explosion imaging (described above in Section~\ref{sec:prog_id}).

\begin{figure}
\plotone{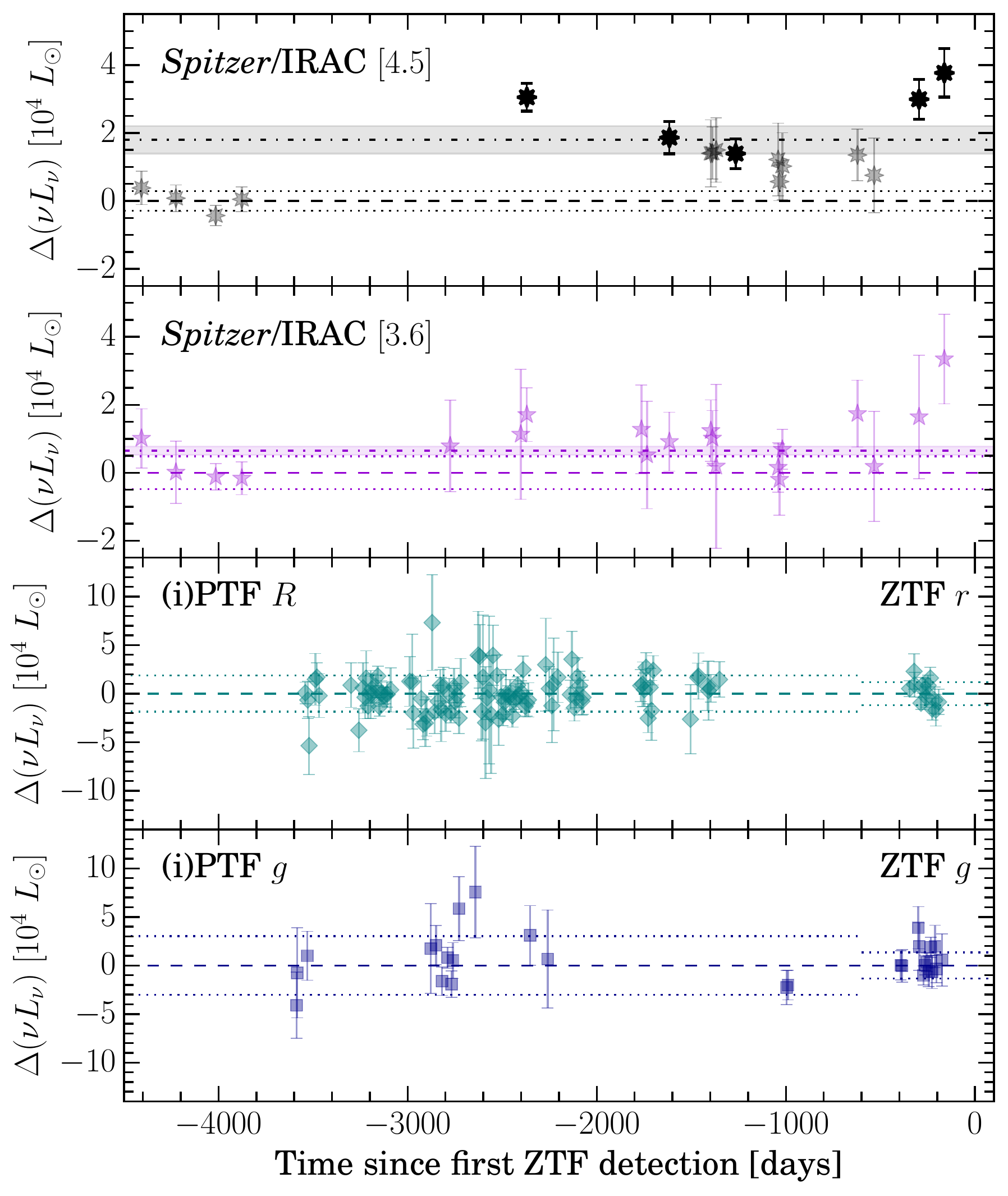}
\caption{\label{fig:pre_explosion_lcs} Constraints on pre-explosion variability based on image subtraction at the location of the transient in the IR with \textit{Spitzer}/IRAC and optical with (i)PTF and ZTF. From top to bottom we show the differential light curves at the [4.5], [3.6], $R$ or $r$, and $g$ bands. The darker, solid black points in the [4.5] light curve highlight individual epochs where we detect variability of the progenitor at $>3\sigma$ significance. The dashed lines indicate the zero level, and dotted lines in each panel show the standard deviations of our measurements for the first four epochs of \textit{Spitzer} imaging, and for the entire sets of (i)PTF and ZTF imaging. The dashed--dotted lines and shaded regions in the top two panels show the ``average'' differential luminosities and corresponding 1$\sigma$ uncertainties measured in the subtraction of the reference Super Mosaic images from our deep stacks of all available pre-explosion \textit{Spitzer}/IRAC [3.6] and [4.5] imaging.
}
\end{figure}

Aperture photometry was performed at the location of M51~OT2019-1 in the difference images, and our resulting differential light curves are shown in Figure~\ref{fig:pre_explosion_lcs}, converted to band luminosities in $\nu L_{\nu}$. During the years 2006--2008 ($\approx 3900$--$4400$~days before discovery), our measurements are consistent with no change compared to the reference level within $|\Delta (\nu L_{\nu})| \lesssim 10^4~L_{\odot}$. In 2012 ($\approx 2400$~days before discovery), we detect a significant, 4$\sigma$-level increase of $\Delta (\nu L_{\nu}) = (3.2 \pm 0.8) \times10^4~L_{\odot}$. Following this, in the time between 2014 and 2017 ($\approx 1600$--$500$ days before discovery), we note a consistently elevated [4.5] flux at the location at $\Delta (\nu L_{\nu}) \approx 1.3\times10^4~L_{\odot}$. In the final two epochs of \textit{Spitzer} coverage, we detect a significant pre-explosion brightening starting sometime between $\approx 500$ and 300~days before discovery, and rising to the level of $\Delta (\nu L_{\nu}) = (3.7 \pm 0.7) \times10^4~L_{\odot}$ at $t = -162.3$~days. At [3.6], the same trends in pre-explosion variability are evident, but at a lower level of significance. In the subtraction of our deep stacks, we detect significant excess flux compared to the reference level of $\Delta (\nu L_{\nu}) = (6.5 \pm 1.3) \times10^3~L_{\odot}$ at [3.6] and $\Delta (\nu L_{\nu}) = (1.8 \pm 0.4) \times10^4~L_{\odot}$ at [4.5], shown as the dashed--dotted lines and shaded regions in Figure~\ref{fig:pre_explosion_lcs}.

While the [4.5] flux of the coincident precursor source measured in our deep pre-explosion stack may contain contamination from nearby sources (Section~\ref{sec:prog_id}), the variable fluxes detected with image subtraction can be unambiguously attributed to a single source. Associating the variable precursor with the OT, these measurements are thus robust lower limits on the average, pre-explosion flux of the progenitor at [3.6] and [4.5] since 2004. We add these constraints on the progenitor SED to Figure~\ref{fig:sed}.

We obtained constraints on pre-explosion optical variability at the location using the available coverage by the Palomar Transient Factory \citep{law09,rau09} and its successor, the intermediate Palomar Transient Factory, (i)PTF \citep{cao16}, and ZTF on P48. For (i)PTF $g$ and Mould-$R$ band data taken between 2009 and 2016, we utilized forced PSF-fitting photometry at the transient location on the reference-subtracted difference images \citep{masci17}. The same procedure as described in Section~\ref{sec:imaging} was used to obtain photometry from ZTF difference images for the entire set of $g$- and $r$-band images covering the site from the publicaly available ZTF-DR1, and the Caltech and Partnership surveys since the start of full operations in 2018 March. To obtain deeper limits, we stacked our measurements from (i)PTF and ZTF within 10\,day windows, and show the resulting differential optical light curves along with the IR measurements in Figure~\ref{fig:pre_explosion_lcs}. We see no evidence for significant optical variability at the location in any of the archival P48 coverage at the level of $|\Delta (\nu L\nu)| \lesssim 2\times10^4$ and $3\times10^4~L_{\odot}$ in the $R$ or $r$ and $g$-bands, respectively, including during the 2012 and 2017--2018 [4.5] brightening episodes. 

\subsection{Photometric properties}\label{sec:phot}
The multiband optical and near-IR light curves of M51~OT2019-1 are shown in Figure~\ref{fig:lcs}. We observe a rise in the optical light curves over a time of $\approx 15$~days after the first detection by ZTF, after which the source exhibits a relatively flat plateau in the $g,r,i$, and $Y$ bands to at least $t=40$~days. By $t\approx50$~days, the light curves begin to decline, particularly noticeable in the $g$ band. In the $r$ band, the source is observed to peak at $M_{r} = -13.0$ (Galactic extinction correction only). The optical colors are remarkably red, with values on the rise of $g - r = 1.3 \pm 0.2$, $g - i = 1.9 \pm 0.2$, and $g - Y = 2.6\pm 0.2$~mag at $t=5.9$~days. On the plateau at $t=25.9$~days, we observe bluer optical colors of $g - r = 0.97 \pm 0.05$, $g - i = 1.53 \pm 0.06$, and $g - Y = 1.8 \pm 0.1$~mag. In the near-IR, the source was detected 6.5~days before the earliest ZTF detection at $K_s = 18.2 \pm 0.3$, before rising to $K_s = 15.66 \pm 0.06$ at $t=28.9$~days.

Given its red optical colors and location in a dark dust lane, it is likely that M51~OT2019-1 is subject to significant extinction. High amounts of optical extinction have also been inferred for similar transients of the ILRT class, but even among these, M51~OT2019-1 is exceptionally red. As shown in Figure~\ref{fig:lcs}, we estimate the excess reddening present in M51~OT2019-1 by comparing its optical light curves to those of the well-studied ILRTs SN\,2008S \citep{botticella09}, NGC\,300~OT2008-1 \citep{humphreys11}, PTF\,10fqs \citep{kasliwal11}, and AT\,2017be \citep{cai18}. After correcting the light curves for Galactic extinction, we require \textit{excess} reddening of $E(B-V) \approx 0.7$, 0.5, 0.4, and 0.6~mag for each of those objects, respectively, to match the optical colors of M51~OT2019-1 near peak on the plateau.

The total extinction to M51~OT2019-1 at a given phase is likely due to a combination of attenuation by foreground dust in the host interstellar medium (ISM) as well as internal extinction by dust in the circumstellar medium (CSM). For NGC\,300~OT2008-1, \citet{humphreys11} estimated the total extinction near the peak of the transient as $E(B-V) \approx 0.4$~mag based on comparing its optical colors to those expected for a temperature $T \approx 7500$~K inferred from the presence of F-type absorption features in the spectrum. A similar argument was made by \citet{smith09} to estimate a total host/CSM extinction to SN\,2008S of $E(B-V) = 0.28$~mag.

As described below in Section~\ref{sec:spectra}, we observe similar absorption features in the spectra of M51~OT2019-1 throughout its evolution, and thus infer a similar temperature for the continuum emission. Our analysis of the SED derived from photometry at several phases in the evolution of the transient is shown in Figure~\ref{fig:sed}. We observed an early brightening at $t=-6$~days in the near-IR $K_s$ band to $\nu L_{\nu} \approx 4\times10^{5}~L_{\odot}$ with a constraining $r$-band nondetection a few days later that suggests the explosion is heavily obscured at early times. At $t=6$~days as the transient rises in the optical, comparing the SED to a reddened blackbody spectrum with $T = 7500$~K provides a good approximation to the data and suggests $E(B-V) = 1.0$~mag and a blackbody radius of $R \approx 2.7\times10^{14}$~cm. Near the optical peak on the plateau at $t=29$~days, the SED appears less reddened, and can be approximated by $T = 7500$~K with $E(B-V) = 0.7$~mag and $R\approx3.8\times10^{14}$~cm. The inferred expansion velocity from the evolution of the blackbody radius of $\approx 550$~km~s$^{-1}$ is notably similar to the observed velocities widths of the H$\alpha$ emission lines ($\approx 400$~km~s$^{-1}$; Section~\ref{sec:spectra}). While some amount of extinction can likely be attributed to foreground host extinction, the variable extinction estimate from evolution of the SED indicates a significant contribution from internal CSM dust. In particular, we suggest the color evolution of the SED from rise to peak may be attributed to the continued destruction of CSM dust as the explosion emerges from the dense obscuring wind of the progenitor. 

The luminosity of the corresponding unreddened blackbody is $L = 8.3\times10^7~L_{\odot}$, which we adopt as a crude estimate of the intrinsic bolometric luminosity of the explosion at peak. In comparison, the most luminous classical novae reached peak luminosities from $\approx 3$--$8\times10^{5}~L_{\odot}$ \citep{gallagher76,gehrz15}, while typical CCSNe reach $5\times10^{8}$ to $\gtrsim 10^{10}~L_{\odot}$ \citep[e.g.,][]{valenti16}. We note, however, the reddened blackbody model significantly overpredicts the observed $Y$-band flux at this epoch. We also see evidence for an IR excess in the near-IR $K_s$-band measurement at $t=29$~days and the later [3.6] and [4.5] measurements from \textit{Spitzer} at $t=75$~days, suggesting emission from dust is an important component of the SED. Finally, in the SED at $t = 41$~days from our \textit{HST} observations (plus symbols in Figure~\ref{fig:sed}), the measurement in F814W lies near the expectation from the reddened blackbody approximation to the $t=29$~day SED, consistent with the flat evolution of the $i^{\prime}$-band light curves between these epochs; however, the F336W and F275W points are significantly below this expectation, suggesting that there may be additional suppression of the flux in the blue and UV (e.g., line blanketing and/or Balmer continuum absorption) and/or that the assumed \citet{fitzpatrick99} $R_V = 3.1$ extinction law for the diffuse Milky Way ISM may not applicable for heavy internal extinction by circumstellar dust, especially at bluer wavelengths. Additional longer wavelength measurements from ongoing monitoring with \textit{Spitzer} and self-consistent modeling of the time evolution of the SED will be necessary to better constrain the intrinsic properties of the explosion and the CSM dust. 

In the right panel of Figure~\ref{fig:lcs}, we compare the $r$-band light curve of M51~OT2019-1 to other hydrogen-rich transients. For the sample of ILRTs, we correct for an estimate of the total extinction based on the SED near peak as described above. Assuming $E(B-V) = 0.7$~mag for M51~OT2019-1, the peak absolute magnitude is $M_{r} \approx -15$. This is notably more luminous than the other proposed members of this class, which fall between $M_{r/R} \approx -13$ to $-14$, and is near the range more typical of Type~II core-collapse SNe. In fact, with our assumed extinction, M51~OT2019-1 is even more luminous than some low-luminosity Type IIP SNe, e.g., SN\,2005cs \citep{pastorello06}. 

\subsection{Spectroscopic properties}\label{sec:spectra}
Our spectroscopic observations of M51~OT2019-1 from $t=4$ to 111\,days are shown in Figure~\ref{fig:spec}. Here, we briefly describe the important spectral properties and features observed in this event, but reserve a detailed analysis of their evolution to future, more comprehensive studies. The spectra are characterized by a very red continuum, and display prominent H$\alpha$ emission along with emission features of the \ion{Ca}{2} IR triplet ($\lambda \lambda~8498, 8542, 8662$) and [\ion{Ca}{2}] ($\lambda \lambda~7291, 7324$). We detect the \ion{Ca}{2} H and K lines ($\lambda \lambda~3968, 3934$), the \ion{Na}{1} D doublet ($\lambda \lambda~5889, 5895$), and a strong \ion{O}{1} blend near 7773~\AA\ in absorption. Superimposed on the broader emission features, we also detect a narrower absorption component for each of the lines of the \ion{Ca}{2} triplet. In the near-IR spectrum at $t=29$~days (top right panel of Figure~\ref{fig:spec}), we see additional recombination lines of H in emission including Pa$\beta$ and weaker Pa$\gamma$. In our latest spectrum at $t=111$~days, the continuum has grown notably redder, and the \ion{Ca}{2} triplet features have transitioned to predominantly narrower emission features.

These features indicate a wide range of densities in the gas producing the spectrum---reminiscent of a wind or ejected envelope with large inhomogeneities. The observed [\ion{Ca}{2}] $\lambda \lambda$ 7291, 7324 lines have a critical density of $n_{\mathrm{cr}} \sim 10^7$~cm$^{-3}$ \citep{ferland89} and are strongly suppressed at higher densities. On the other hand, the \ion{O}{1} $\lambda$7773 transition, a quintet frequently observed in transients, is normally populated by collisional transfer from the \ion{O}{1} 3p $^3$P level, which is itself strongly
excited by a wavelength coincidence between the \ion{H}{1} Ly\,$\beta$ and \ion{O}{1} $\lambda$1026 transitions \citep{bowen47}. The triplet-to-quintet collisional transfer is only effective at densities $n\gtrsim10^{11}$~cm$^{-3}$ \citep{williams12}, thus demonstrating a wide range of densities in the gas ejected by the outburst. 

Furthermore, as discussed by \citet{humphreys11} for NGC\,300~OT2008-1, the \ion{Ca}{2} and [\ion{Ca}{2}] emission lines are sensitive to conditions in the ejecta and clearly evolve with time. The triplet emission lines are formed in the ejecta by radiative de-excitation of electrons in upper levels of the \ion{Ca}{2} H and K absorption transitions, leaving the electrons in the upper levels of the [\ion{Ca}{2}] lines \citep{ferland89}. Typically, these electrons are collisionally de-excited to the ground state, unless the densities are low enough. Following \citet{humphreys11,humphreys13}, we can estimate the fraction of photons that are radiatively de-excited from the ratio of the equivalent widths of the [\ion{Ca}{2}] lines to the \ion{Ca}{2} triplet, times the ratio of the expected fluxes for the continuum emission at the corresponding wavelengths of 7300 and 8600~\AA. Assuming a $7500$~K blackbody for the continuum, we estimate photon fractions of $\approx 0.25$ and $0.3$ in our $t=5$ and $44$~day spectra, respectively, but find a lower value of $\approx 0.1$ at $t=111$~days. This is remarkably similar to the evolution for NGC\,300~OT2008-1. As suggested by \citet{humphreys11}, the decline in the photon ratio at later times may indicate an increase in the density where the [\ion{Ca}{2}] lines are formed in the ejecta.

The H$\alpha$ velocity profiles (bottom right panel of Figure~\ref{fig:spec}) appear symmetric about their peaks, which are consistent with zero velocity in the rest frame of M51. The profiles are characterized by a FWHM velocity of $\approx 400$~km~s$^{-1}$, much lower than typical velocities observed in CCSNe of $\sim 10,000$~km~s$^{-1}$, with broader Thomson-scattering wings extending to $\approx 2000$~km~s$^{-1}$, similar to those seen in NGC\,300~OT2008-1 \citep{humphreys11}. The $t = 5$~days GMOS spectrum shows an apparent absorption feature near $\approx -180$~km~s$^{-1}$, but we attribute this to a data reduction artifact from oversubtraction of unrelated, background H$\alpha$ emission and difficultly in finding a suitably clean region for background subtraction along the slit. We do not see any evidence for significant evolution in the line profile shape or width for the duration of the observations presented here. 

The observed absorption spectrum is reminiscent of an F-supergiant. This is expected for an eruption that produces an extended, optically thick wind \citep{davidson87}, and is seen in LBVs/S~Doradus variables in their cool, outburst state at maximum \citep{humphreys94}, some SN impostors and LBV giant eruptions \citep[e.g.][]{smith11}, and is very similar to that of NGC\,300~OT2008-1 \citep{bond09,humphreys11} and SN\,2008S \citep{smith09}. This suggests a temperature of $\approx 7500$~K. The red continuum is therefore suggestive of significant extinction, and we apply this temperature estimate to our SED analysis in Section~\ref{sec:phot} to obtain $E(B-V) = 0.7$~mag near the peak at $t=29$~days. Overall, the spectral features described above are generally similar to some previously observed SN impostors and ILRTs.

\section{Discussion and Conclusions}\label{sec:discussion}
From our early observations of M51~OT2019-1, we find that the transient is characterized by a $\approx15$\,day rise in the optical to an observed plateau luminosity of $M_r = -13$ (Galactic extinction correction only). Its spectrum shows strong H$\alpha$ emission with FWHM velocities of $\approx 400$~km~s$^{-1}$. These basic properties are similar to those of multiple classes of transients with luminosities intermediate between those of novae and supernovae, including ILRTs, LRNe, and giant eruptions of LBVs. The photometric evolution of M51~OT2019-1 has transitioned smoothly from the rise to a plateau and subsequent smooth decline after $t\approx 50$~days, consistent with ILRTs and also similar to some SN impostors. It is more dissimilar to LRNe whose light curves are typically irregular and multi-peaked \citep{sparks08,smith16,blagorodnova17} and some other LBV eruptions that may show multiple outbursts or erratic variability \citep{pastorello10,smith10}. Our optical spectra show several additional features, including \ion{Ca}{2} and [\ion{Ca}{2}] in emission, F-type absorption features indicating the light is escaping an optically thick wind at $\approx 7500$~K, and a red continuum indicative of significant extinction. These features are characteristic of ILRTs, but may also be seen in LRNe and giant LBV eruptions. For example, while some LBV eruptions show bluer continua and H lines with strong P Cygni profiles (e.g., NGC\,3432~OT; \citealp{pastorello10}), others are indistinguishable from ILRTs \citep{smith10, smith11, rest12, prieto14}. We identify a likely progenitor star in archival \textit{Spitzer} imaging with $M_{[4.5]} = -12.2$~mag and $[3.6] - [4.5] > 0.74$~mag, but there is no detected optical counterpart in archival \textit{HST} imaging. This is reminiscent of properties of the obscured progenitors of SN\,2008S and NGC\,300~OT2008-1. We thus suggest M51~OT2019-1 is an ILRT, but with only weak archival constraints at longer wavelengths, we cannot rule out that the progenitor of M51~OT2019-1 is more luminous and massive than the $9$--$15~M_{\odot}$ stars inferred for the ILRT prototypes. 

The nature of ILRTs, the mechanism behind their outbursts, and their relation to other impostors and LBV-related transients are debated. One proposed physical scenario involves a weak explosion, possibly the electron-capture-induced collapse of an extreme asymptotic giant branch (AGB) star. In this scenario, an initial flash rapidly destroys the enshrouding circumstellar dust, which later reforms and re-obscures the optical transient \citep{thompson09,kochanek11,szczygiel12}. Thanks to the early discovery of M51~2019OT-1, we find evidence for continued dust destruction during the rise of the transient as our estimate of the reddening evolves from $E(B-V) = 1.0$ to $0.7$~mag between $t=6$ and $29$~days, posing a challenge to such interpretations for this event. Alternatively, \citet{humphreys11} suggested that the progenitors of NGC\,300~OT2008-1 and, by extension, SN\,2008S may have been post-AGB stars in transition to warmer temperatures, thus near their Eddington limits and subject to a range of instabilities. Adopting $E(B-V) = 0.7$~mag at peak for M51~OT2019-1, the intrinsic luminosity is $M_{r} = -15$~mag, higher than previously observed ILRTs, though not unusual for some LBV-related SN impostors \citep{smith11}, possibly suggesting that these events may represent a continuum of related, explosive phenomena arising from evolved progenitors spanning a wide range of masses. Finally, the observed IR variability of the likely progenitor may be consistent with a long-period ($\sim 2000$~days), thermally pulsing super-AGB (see the candidate super-AGB MSX~SMC~055; \citealp{groenewegen09}), or may hint at eruptive, self-obscuring episodes occurring in the years before the explosion. A complete picture of the event awaits continued monitoring, including ongoing \textit{Spitzer} observations, to characterize the full SED, dust properties, and energetics. Longer-term, mid-IR spectroscopic observations with the \textit{James Webb Space Telescope} will disentangle the chemistry, and late-time imaging after the explosion fades away will provide definitive evidence on whether the explosion was terminal. 

\acknowledgments
We thank J.\ Prieto and E.\ Ofek for valuable discussions in revising this work. We also thank the anonymous referee for their helpful comments, which improved the manuscript.

This material is based upon work supported by the National Science Foundation Graduate Research Fellowship under grant No.\ DGE-1144469. H.E.B. acknowledges support from program numbers GO-14258 and AR-15005, provided by NASA through grants from the Space Telescope Science Institute, which is operated by the Association of Universities for Research in Astronomy, Incorporated, under NASA contract NAS5-26555. This work is part of the research programme VENI, with project number 016.192.277, which is (partly) financed by the Netherlands Organisation for Scientific Research (NWO). A.G.-Y. is supported by the EU via ERC grant No.\ 725161, the ISF, the BSF Transformative program and by a Kimmel award. R.D.G. was supported by NASA and the United States Air Force. Based on observations obtained with the Samuel Oschin Telescope 48\,inch and the 60\,inch Telescope at the Palomar Observatory as part of the Zwicky Transient Facility project. ZTF is supported by the National Science Foundation under grant No.\ AST-1440341 and a collaboration including Caltech, IPAC, the Weizmann Institute for Science, the Oskar Klein Center at Stockholm University, the University of Maryland, the University of Washington, Deutsches Elektronen-Synchrotron and Humboldt University, Los Alamos National Laboratories, the TANGO Consortium of Taiwan, the University of Wisconsin at Milwaukee, and Lawrence Berkeley National Laboratories. Operations are conducted by COO, IPAC, and UW. This work was supported by the GROWTH project funded by the National Science Foundation under grant No.\ 1545949. SED Machine is based upon work supported by the National Science Foundation under Grant No.\ 1106171. The work presented here is based on observations obtained with the Apache Point Observatory 3.5\,m telescope, which is owned and operated by the Astrophysical Research Consortium. We thank the Apache Point Observatory Observing Specialists for their assistance during the observations. The data presented here were obtained in part with ALFOSC, which is provided by the Instituto de Astrofisica de Andalucia (IAA) under a joint agreement with the University of Copenhagen and NOTSA. Based on observations obtained at the Gemini Observatory acquired through the Gemini Observatory Archive and processed using the Gemini IRAF package, which is operated by the Association of Universities for Research in Astronomy, Inc., under a cooperative agreement with the NSF on behalf of the Gemini partnership: the National Science Foundation (United States), National Research Council (Canada), CONICYT (Chile), Ministerio de Ciencia, Tecnolog\'{i}a e Innovaci\'{o}n Productiva (Argentina), Minist\'{e}rio da Ci\^{e}ncia, Tecnologia e Inova\c{c}\~{a}o (Brazil), and Korea Astronomy and Space Science Institute (Republic of Korea). Some of the data presented herein were obtained at the W.\ M.\ Keck Observatory, which is operated as a scientific partnership among the California Institute of Technology, the University of California and the National Aeronautics and Space Administration. The Observatory was made possible by the generous financial support of the W.\ M.\ Keck Foundation. The authors wish to recognize and acknowledge the very significant cultural role and reverence that the summit of Maunakea has always had within the indigenous Hawaiian community.  We are most fortunate to have the opportunity to conduct observations from this mountain. This work is based in part on observations made with the {\it Spitzer Space Telescope}, which is operated by the Jet Propulsion Laboratory, California Institute of Technology under a contract with NASA. Based on observations made with the NASA/ESA {\it Hubble Space Telescope}, obtained from the Data Archive at the Space Telescope Science Institute, which is operated by the Association of Universities for Research in Astronomy, Inc., under NASA contract NAS 5-26555. Some of these observations are associated with program \#15675. This publication makes use of data products from the \textit{Wide-field Infrared Survey Explorer}, which is a joint project of the University of California, Los Angeles, and the Jet Propulsion Laboratory/California Institute of Technology, funded by the National Aeronautics and Space Administration. This research has made use of the NASA/IPAC Extragalactic Database (NED), which is operated by the Jet Propulsion Laboratory, California Institute of Technology, under contract with the National Aeronautics and Space Administration.

\facilities{PO:1.2m (ZTF), PO:1.5m (SEDM), Hale (DBSP, WIRC, TripleSpec), NOT (ALFOSC), Gemini:Gillett (GMOS), Keck:I (LRIS), Keck:II (NIRC2), ARC (NICFPS, ARCTIC), \textit{Spitzer} (IRAC), \textit{HST} (ACS, WFC3), \textit{WISE}.}

\software{AstroDrizzle \citep[][\url{http://drizzlepac.stsci.edu}]{hack12}, DOLPHOT \citep{dolphin00,dolphin16}, IRAF \citep{tody86,tody93}, Gemini IRAF package (\url{http://www.gemini.edu/sciops/data-and-results/processing-software}), PyRAF (\url{http://www.stsci.edu/institute/software_hardware/pyraf}), pyraf-dbsp \citep[][\url{https://github.com/ebellm/pyraf-dbsp}]{bellm16}, LPipe \citep[][\url{http://www.astro.caltech.edu/~dperley/programs/lpipe.html}]{perley19}.}

\bibliographystyle{yahapj}
\bibliography{M51OT}

%\appendix
%\section{appendix section}

\end{document}